\documentclass{esannV2}
\usepackage[dvips]{graphicx}
\usepackage[latin1]{inputenc}
\usepackage{amssymb,amsmath,array}

\begin{document}
\title{Calliope - A Polyphonic Music Transformer}

\author{Andrea Valenti, Stefano Berti and Davide Bacciu
%
\thanks{This research was partially supported by H2020 TAILOR (GA 952215)}
%
\vspace{.3cm}\\
%
University of Pisa - Dept. of Computer Science \\
Largo B. Pontecorvo, 3, 56127 Pisa - Italy
%
}

\maketitle

\begin{abstract}
The polyphonic nature of music makes the application of deep learning to music modelling a challenging task. On the other hand, the Transformer architecture seems to be a good fit for this kind of data. In this work,  we present Calliope, a novel autoencoder model based on Transformers for the efficient modelling of multi-track sequences of polyphonic music. The experiments show that our model is able to improve the state of the art on musical sequence reconstruction and generation, with remarkably  good results especially  on  long sequences. 
\end{abstract}

\section{Introduction}

Language and music are often considered to be characterized by an high degree of resemblance. 
For instance, both can be represented by sequences of symbols: words for text, notes for music.
Those symbolic sequences present well-defined structures that are often articulated at different layers of abstraction.
However, music poses additional challenges over text: unlike the latter, a musical sequences can contain chords instead of simple notes, with more than one symbol being simultaneously played.
A song is also often the combination of a series of instruments, each defining a different sub-sequence of note symbols played at the same time.
The polyphonic nature of music, both at note and track level, poses several challenges to recent deep learning models, making their application to music modelling a challenging task.

Notable models that tackle the problem of automatic music generation are the MusicVAE \cite{roberts2018hierarchical} and MusAE \cite{valenti2020musae}, both of which use a LSTM-based architecture to generate sequences of monophonic music.
MuseGAN \cite{dong2018musegan} combines generative adversarial networks (GAN) \cite{goodfellow2014generative} with a convolutional architecture to model multi-track MIDI polyphonic music. 
However, results are limited to relatively short (4-bar) sequences. One of the first attempts to use Transformer architecture \cite{vaswani2017attention} for music is the Music Transformer \cite{huang2020pop}. The new architecture allows the model to generate longer single-track sequences of piano music. Transformer Autoencoder \cite{choi2020encoding} use the Transformer to generate new songs in the style of a given performance. Despite the promising results, these models are only evaluated on simple datasets, only containing songs played on a piano with homogeneous stylistic characteristics.

In this work, we present Calliope, a novel deep learning model for the automatic generation of music. 
To the best of our knowledge, this is the first Transformer-based architecture for challenging multi-track polyphonic music reconstruction and generation, comprising songs from multiple artists and genres.

\section{Calliope: Polyphonic Music with Transformers AAE}
Calliope's architecture fits into the framework of Adversarial Autoencoders (AAE) \cite{makhzani2015adversarial}. That is a variational autoencoder \cite{kingma2013auto} augmented with an additional network (the \textit{Discriminator}) that performs the variational approximation of the posterior in an adversarial way. The model is trained to maximize the Evidence Lower BOund (ELBO): given a data sample $x$, the ELBO is defined as
\begin{equation}
    \label{eq_elbo}
    \mathcal{L}(x, \theta, \phi) = \mathbb{E}_{q_{\phi}(z|x)}\left[ p_\theta(x|z) \right] - \beta D_{KL}\left( q_\phi(z|x) || p(z) \right)
\end{equation}
where $p_\theta(x|z)$ is the decoder's distribution, $q_\phi(z|x)$ is the encoder's distribution and $p(z)$ is the chosen prior distribution of the latents, $\beta$ is an hyperparameter that controls the amount of desired regularization.
Calliope's training algorithm comprises two main phases: the reconstruction phase and regularization phase.
In the reconstruction phase, only the first term of Eq.\ref{eq_elbo} is optimized. A batch of songs is fed in input to the encoder to get their respective latent encodings, which are then decoded back into the original songs. The model is trained to have the reconstructed songs to be as close as possible to the originals.
The regularization phase optimizes the second term of Eq.\ref{eq_elbo}. The KL divergence is approximated with the aid of an additional discriminator network $d_\psi(z)$ using the density-ratio trick \cite{mescheder2017adversarial}.
This induces a min-max adversarial game between the encoder $q_\phi(z|x)$ and the discriminator $d_\psi(z)$ similar to the one of GANs:
\begin{equation}
    \min_\phi\max_\psi \mathbb{E}_{q_\phi(z|x)}[ \log d_\psi(z)] + \mathbb{E}_{p(z)}[\log(1-d_\psi(x))].\nonumber
\end{equation}
While the \textit{Discriminator} is implemented as a multi-layer perceptron, the encoder and the decoders use purposely developed and tailored Transformer architectures \cite{vaswani2017attention}, described in the following sections.

\subsection{Data Representation}
A note is represented as a tuple $(time, pitch, duration)$, where \textit{time} denotes the note onset time (measured as the offset, in number of timesteps, from the start of the measure), \textit{pitch} is the actual pitch played by the note, and \textit{duration} is the number of timesteps during which the note keeps playing. Each element of this tuple is represented by a symbolic token, taken from the following vocabulary: 128 pitch values (tokens from 0 to 127),  96 time values  (tokens from 128 to 223), 96 duration values (tokens from 224 to 319), pad (token 320), start-of-song and end-of-song (tokens 321 and 322). 
Before feeding the sequence in input to the encoder, the tokens are embedded in a vector of $N_x$ element, which is learned during training along with the other model parameters.

\subsection{The Calliope Encoder}
The encoder compresses the song $x$, into a single latent code $z$ of dimension $N_z$. The encoder's forward pass is the following:
\begin{align}
h_{i, t} &\xleftarrow{} \mathrm{Encoder}(x_{i, t})\nonumber\\
z_{i, t} &\xleftarrow{} \mathrm{Comp}(h_{i, t})\nonumber\\
z_{i} &\xleftarrow{} \mathrm{BarCompressor}([z_{i,t_1} ...  z_{i, t_M}])\nonumber \\
z &\xleftarrow{} \mathrm{SongCompressor}([z_{1} ... z_{N}])\nonumber
\end{align}
where $N$ is the song length, $M$ the number of tracks, $t$ is the track index and $i$ is the measure index.
First, $x$ is split along the time axis into a series 1-measure sequences $x_i$. Then, each $x_i$ is further split into each track that composes a song $x_{i, t}$. These single-track, single-measure sequences are finally encoded separately via the  \textit{Encoder} module, which is implemented as a Transformer with the addition of Relative Positional Encoding in the Self-Attention Layer \cite{dai2019transformer}. A different Encoder for each instrument is used.
Since the \textit{Encoder} preserves the original dimensions of the input, we use the \text{Comp} module to compress $h_{i, t}$ along the time dimension in order to obtain a single $z_{i, t}$ for the measure. The $z_{i, t}$ of the individual tracks are then merged together: the \textit{BarCompressor} concats the $z_{i, t}$ along the track dimension and passes them through a linear layer to get a single code for the multi-track measure $z_i$. Each $z_i$ is further concatenated with the other codes 
of the sequence along the time dimension, and the resulting tensor is the fed to the \textit{SongCompressor}, consisting in a combination of a linear and layernorm\cite{vaswani2017attention} layers, in order to get the final latent representation of the song $z$.

\subsection{The Calliope Decoder}
The decoder's task is specular to the encoder's. It processes a single latent vector $z$ in order to generate a multi-track polyphonic musical sequence $x$:
\begin{align}
[z_{1}, ... z_{i}, ... z_{N}] &\xleftarrow{} \mathrm{SongDecompressor}(z)\nonumber\\
[z_{i,t_1} ...  z_{i, t_M}] &\xleftarrow{} \mathrm{BarDecompressor}(z_{i})\nonumber\\
h_{i, t} &\xleftarrow{} \mathrm{Decomp}(z_{i, t})\nonumber\\
x_{i, t} &\xleftarrow{} \mathrm{Decoder}(h_{i, t})\nonumber
\end{align}
where $N$, $i$ and $t$ are, again, the song length, the measure index and the track index respectively. First, $z$ is sent into a \textit{SongDecompressor} module that repeats $z$ for the length of the sequence (in measures) and projects it through a linear layer to get a series of $z_i$, each of them responsible for encoding a single measure. The single-measure, single-track codes $z_{i,t}$ are computed from the $z_i$ by the \textit{BarDecompressor} module, that splits the $z_i$ into equal parts (one for each track) and projects each part into the appropriate shape using a linear layer. The \textit{Decomp} module restores the time dimension of the $z_{i,t}$, making them ready to be processed by the \textit{Decoder}, which is again a Transformer. The output of the \textit{Decomp} module is used as memory into the source attention layer of the decoder to guide the generation of the final song.

\section{Experiments}
We performed an experimental analysis to asses Calliope's capabilities using the Lakh MIDI Dataset \cite{raffel2016learning}, a collection of about 100k MIDI songs of various artists and genres. The model is trained using the Adam optimizer \cite{kingma2014adam} with a learning rate of $10^{-4}$. We trained three models with sequence length of 1-bar (96 timesteps), 2-bars (192 timesteps) and 16-bars (1536 timesteps), respectively. The embeddings dimension is 256. The \textit{Encoder} and \textit{Decoder} modules have 6 Transformer layers of size 512. We use 4 heads in the multi-head attention layer. The dimension of the latent space is 256. The batch size is 20 for the 1-bar and 2-bars models and 2 for the 16-bars model, due to memory limitation of the training hardware. The dataset has been split into 70\% songs for training, 10\% for validation and the remaining 20\% for test. For decoding we use a  scheduled sampling strategy  \cite{mihaylova2019scheduled} with $K$=1 and teacher forcing probability of $0.5$. After 50k training steps for the 1-bar and 2-bars models and 25k steps for the 16-bars model, the value of $\beta$ is gradually annealed from $0$ to $0.1$. In the following experiments, we chose the prior distribution $p(z)$ to be an isotropic standard Gaussian $\mathcal{N}(0, I)$.
The source code can be found online\footnote{\texttt{https://www.github.com/StefanoBerti/MusAE}}.
Finally, recognizing that ``writing about music is like dancing about architecture", we provide additional samples of the generated songs\footnote{ \texttt{https://stefanoberti.github.io/musae}}.

\begin{table}
    \begin{center}
    \makebox[\textwidth][c]{
        \begin{tabular}{|l|c|c|c|c|}
            \hline
            \textbf{} & \textbf{Calliope} & \textbf{MusAE} & \textbf{MusicVAE} & \textbf{MusicVAE H}\\
            \hline
                16-bar drums (seq) & \textbf{0.961} & 0.773 & 0.641 & 0.895\\
                16-bar melody (seq)&  \textbf{0.890} & 0.710 & 0.660 & 0.760\\
            \hline
                2-bar drums (seq) & 0.989 & \textbf{0.999} & 0.917 & -\\ 
                2-bar melody (seq) & 0.967 & \textbf{0.991} & 0.951 & -\\ 
            \hline
                16-bar drums (next) & \textbf{0.952} & - & 0.884 & 0.928 \\
                16-bar melody (next) & \textbf{0.923} & - & 0.919 & 0.919\\
            \hline
                2-bar drums (next)& \textbf{0.995} & - & - & 0.979\\
                2-bar melody (next)& \textbf{0.991} & - & - & 0.986\\
            \hline
        \end{tabular}
    }
        \caption{Reconstruction accuracy for Callipoe, MusAE, flat and hierarchical versions of Music VAE. (seq) is whole sequence reconstruction accuracy, while (next) is the next step prediction accuracy.}
        \label{tab_reconstruction}    
    \end{center}
\end{table}

\subsection{Sequence Reconstruction}
The first set of experiments explores the auto-encoding abilities of the model.
The reconstruction accuracy of our model is compared with the MusicVAE \cite{roberts2018hierarchical} and MusAE \cite{valenti2020musae} models. Table \ref{tab_reconstruction} reports the results on the test set. In general, our model exhibits a higher reconstruction accuracy than its competitors, with only a slightly lower reconstruction accuracy than MusAE on shorts sequences. This can be explained by the fact that we are now dealing with polyphonic tracks, while the other two models are capable of processing monophonic music only. The accuracy is significantly higher in the case of 16-bar sequences, showing that the new architecture is particularly suited for the processing of long-range dependencies in the input.
From a qualitative point of view, the discrepancies between the reconstructions and the originals still preserve the tonal characteristics of the songs and generally do not negatively affect the quality of the reconstructed sequence in a significant way.


\begin{table}
\makebox[\textwidth][c]{
    \begin{tabular}{|c|cccc|ccc|}
\hline
      & \multicolumn{4}{c|}{\textbf{EB}} & \multicolumn{3}{c|}{\textbf{UPC}}\\
      
 & \textbf{B}  & \textbf{D} & \textbf{G/P} & \textbf{S} & \textbf{B}  & \textbf{G/P} & \textbf{S}\\
 \hline
jamming & 6.59 & 2.33 & 20.45 & 6.10 & \textbf{1.53} & 3.91 & 4.09 \\
composer & 0.01 & 28.9 & 1.35 & 0.01 & 2.51 & 4.55 & 5.19 \\
hybrid & 2.14 & 29.7 & 14.75 & 6.04 & 2.35 & 5.11 & 5.24 \\ \hline
Calliope & \textbf{0.0} & \textbf{0.0} & \textbf{0.0} & \textbf{0.0} & 2.08 & \textbf{3.87} & \textbf{2.52} \\
\hline
\end{tabular}
}
\end{table}
\begin{table}
\makebox[\textwidth][c]{
    \begin{tabular}{|c|ccc|c|}
\hline
       & \multicolumn{3}{c|}{\textbf{QN}} & \multicolumn{1}{c|}{\textbf{DP}}\\

  & \textbf{B} & \textbf{G/P} & \textbf{S} & \textbf{D}\\
 \hline
jamming & 71.5 & 59.6 & 63.1 & 93.2\\
composer & 49.5 & 48.65 & 52.5 & 75.3\\
hybrid & 44.6 & 44.35 & 52.0 & 71.3\\ \hline
Calliope & \textbf{99.0} & \textbf{96.15} & \textbf{96.21} & \textbf{94.84}\\
\hline
\end{tabular}
}
  \caption{Comparison of generation metrics between the jamming, composer and hybrid models of MuseGAN and Calliope.}
\label{tab_metrics}
\end{table}

\subsection{Song Generation}
The second aspect that we wish to explore is the ability of our model to generate from scratch new musical sequences by decoding new latent codes directly sampled from the prior. Therefore, we decode a set of 20000 latent codes sampled directly from the prior and compute a set of metrics on them: 1) EB: ratio of empty measures. A high EB could indicate ``holes" into the latent space and, in general, can be a symptom of problems during the training procedure. 2) UPC: number of used pitch classes per bar. Lower UPC means the the model is able to choose (and stick to) a specific tonality, reducing the probability of generating dissonant notes and chords. 3) QN: ratio of ``qualified" notes. A note no shorter than three time steps is a qualified note. QN shows whether the music is overly-fragmented with poor or absent structure. 4) DP: drum patterns. Ratio of notes in 8 or 16-beat patterns, common ones for songs in 4/4 time. High DP shows that the model knows how to emphasize the correct timesteps of generated measures. Results are reported in Table \ref{tab_metrics}, compared to the MuseGAN \cite{dong2018musegan} model. Our model scores very well on all the considered metrics, showing that the generated sequences have everything it takes to be appealing for the human listener.

\section{Conclusion and Future Work}
In this paper we presented Calliope, a novel Transformer-based architecture for the efficient modelling of multi-track polyphonic music. The experiments show that our model is able to improve the state of the art on musical sequence reconstruction and generation, with remarkably good results especially on long sequences. The generated songs are musically meaningful and have similar characteristics of human-generated music.
In the future, we plan to increase the number of tracks that the model is able to process. Furthermore, we plan to add more controllable musical properties such as notes velocity, tempo and key changes.


\begin{footnotesize}

\bibliographystyle{unsrt}
\bibliography{esannV2}

\end{footnotesize}


\end{document}